\documentclass[aps,groupedaddress,prb,notitlepage]{revtex4-1}

\usepackage{amsmath}
\usepackage{amssymb}
\usepackage{graphicx}
\usepackage{bm}
\usepackage{epstopdf}
\newcommand{\zbp}{z_\mathrm{BP}}

\begin{document}
\bibliographystyle{apsrev4-1}
\title{Pinning of a Bloch point by an atomic lattice}
\author{Se Kwon Kim}
\author{Oleg Tchernyshyov}
\affiliation{Department of Physics and Astronomy, Johns Hopkins University, 3400 N. Charles St., Baltimore, Maryland, 21218}
\date{\today}

\begin{abstract}
Bloch points are magnetic topological defects. The discrete nature of a magnetic lattice creates a periodic potential that can pin a Bloch point. The pinning force is of the order of the exchange constant, a few piconewtons in a typical ferromagnet (permalloy). A domain wall containing a Bloch point can have a sizable depinning field in the tens of oersted.
\end{abstract}

\maketitle

\section{Introduction}
\label{sec:introduction}

\begin{figure}
\includegraphics[width=0.2\columnwidth]{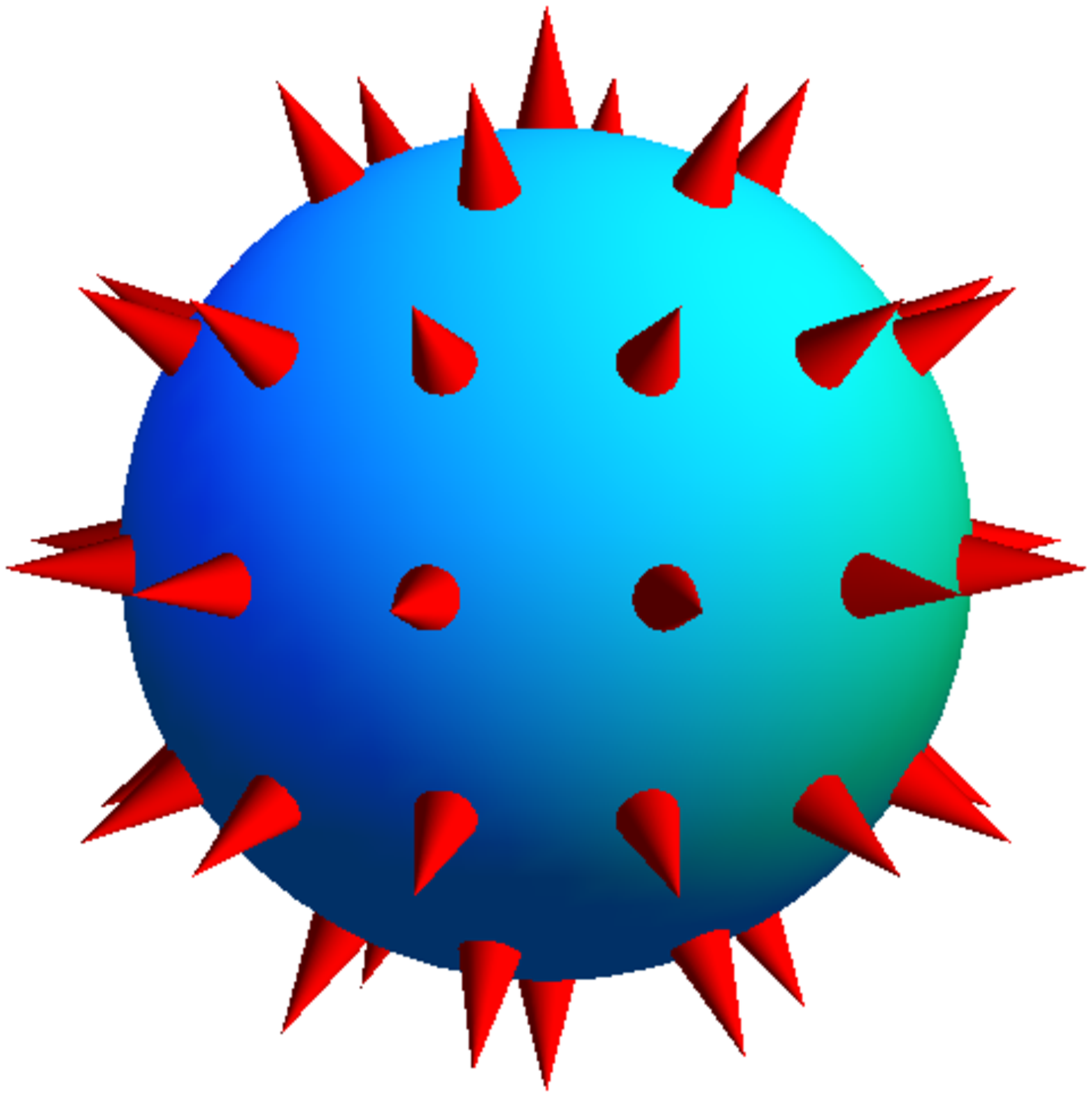} 
\hskip 0.5cm
\includegraphics[width=0.2\columnwidth]{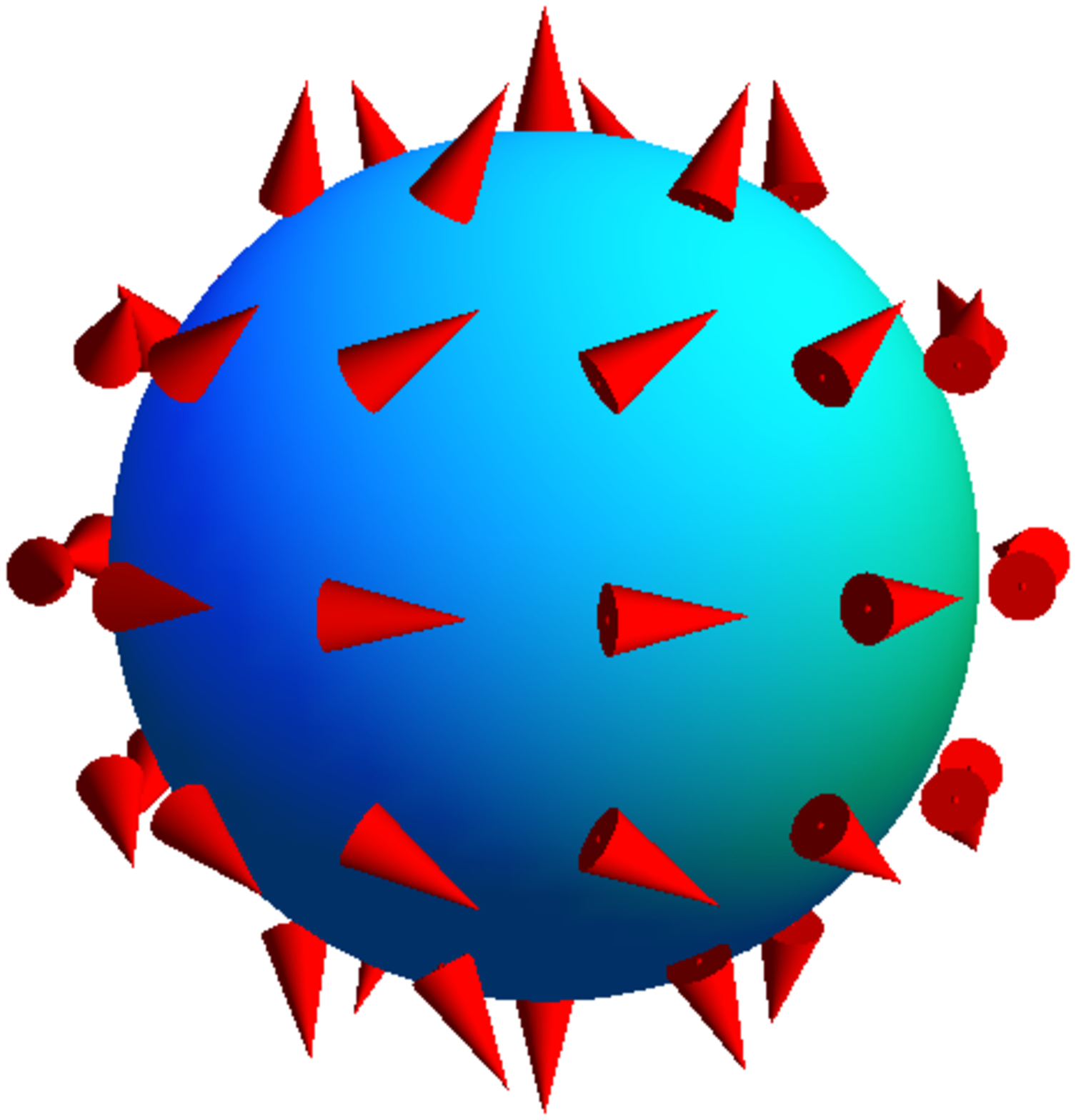} 
\hskip 0.5cm
\includegraphics[width=0.2\columnwidth]{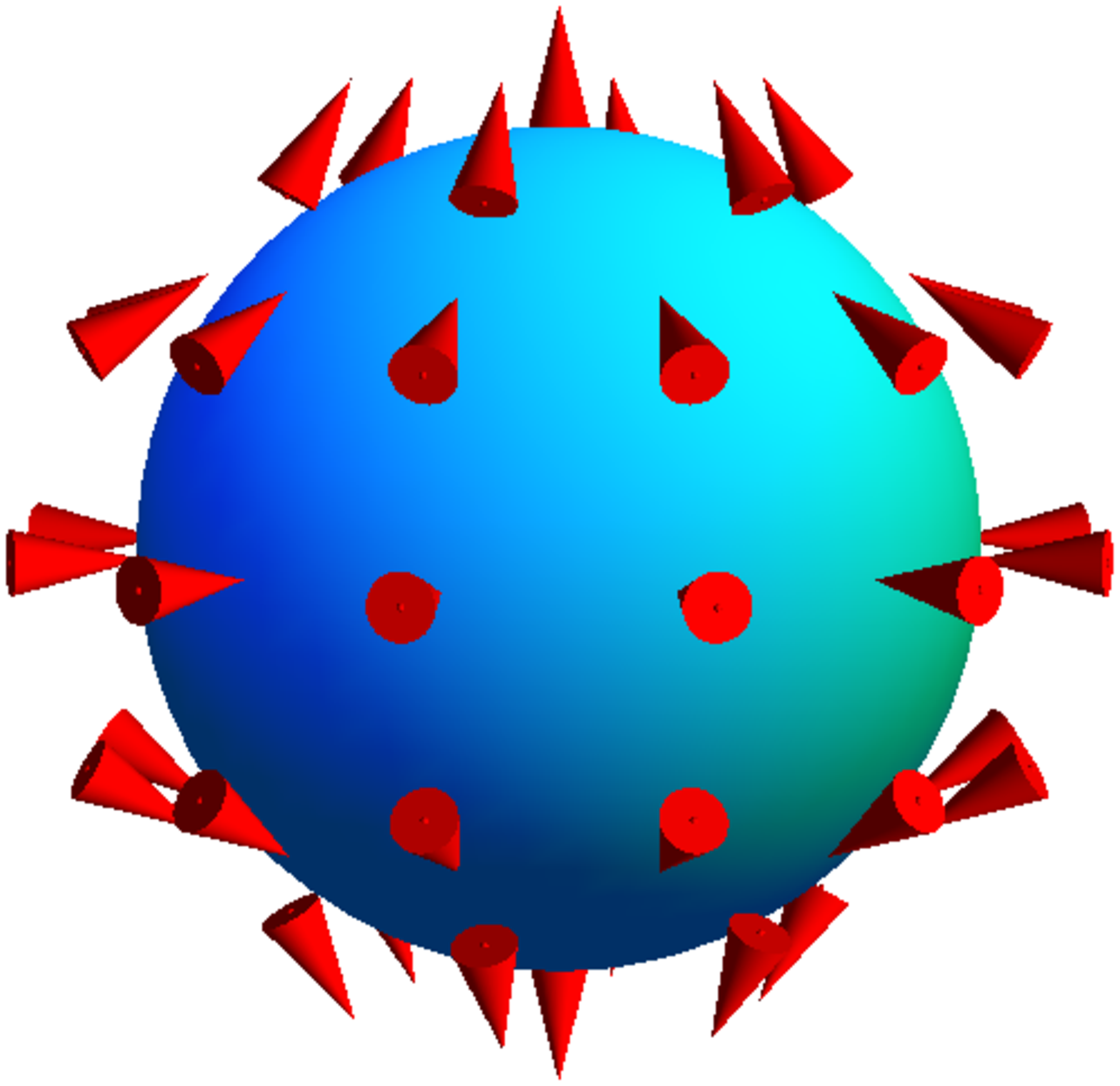} 
\\
\includegraphics[width=0.2\columnwidth]{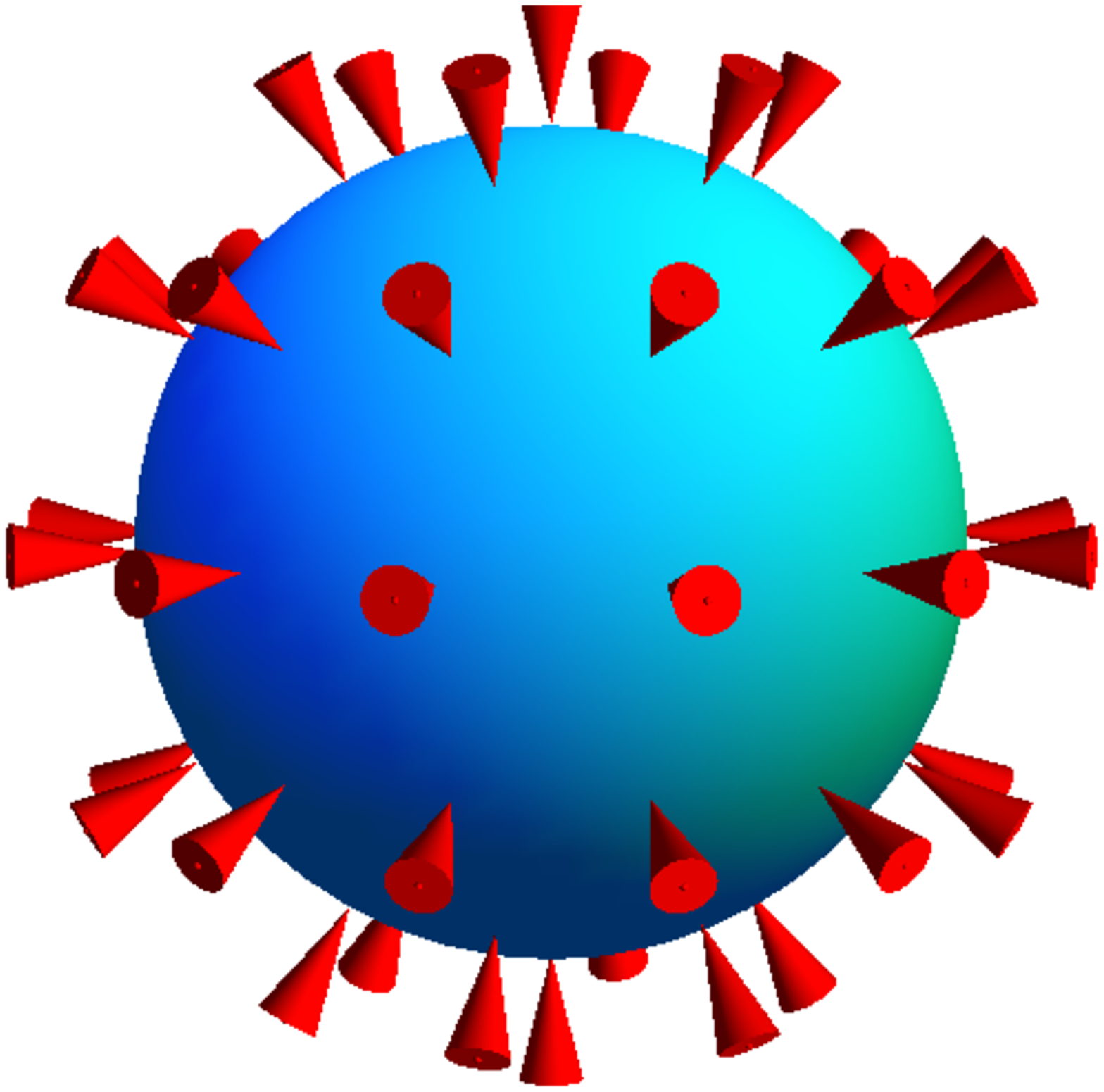} 
\hskip 0.5cm
\includegraphics[width=0.2\columnwidth]{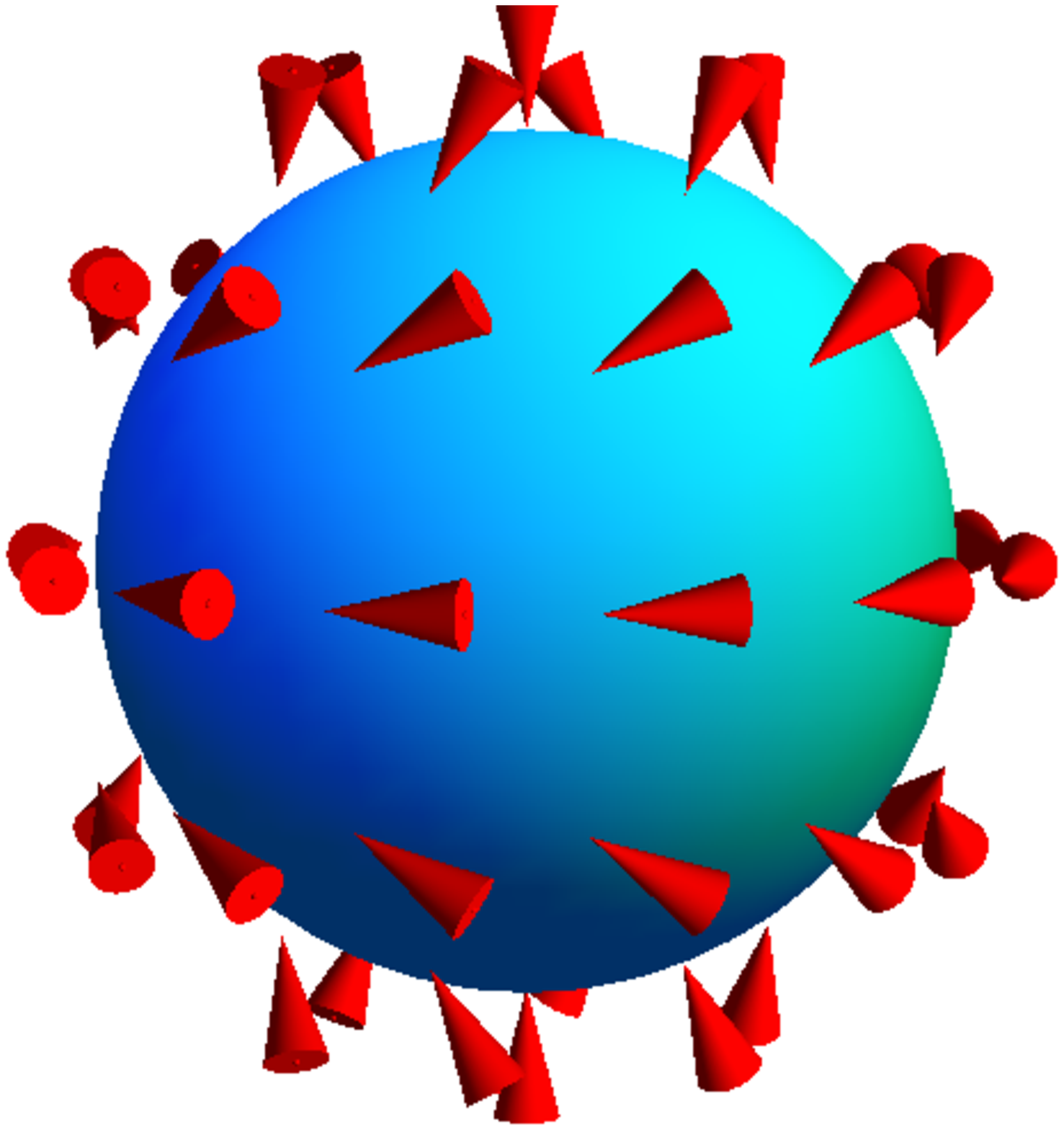} 
\hskip 0.5cm
\includegraphics[width=0.2\columnwidth]{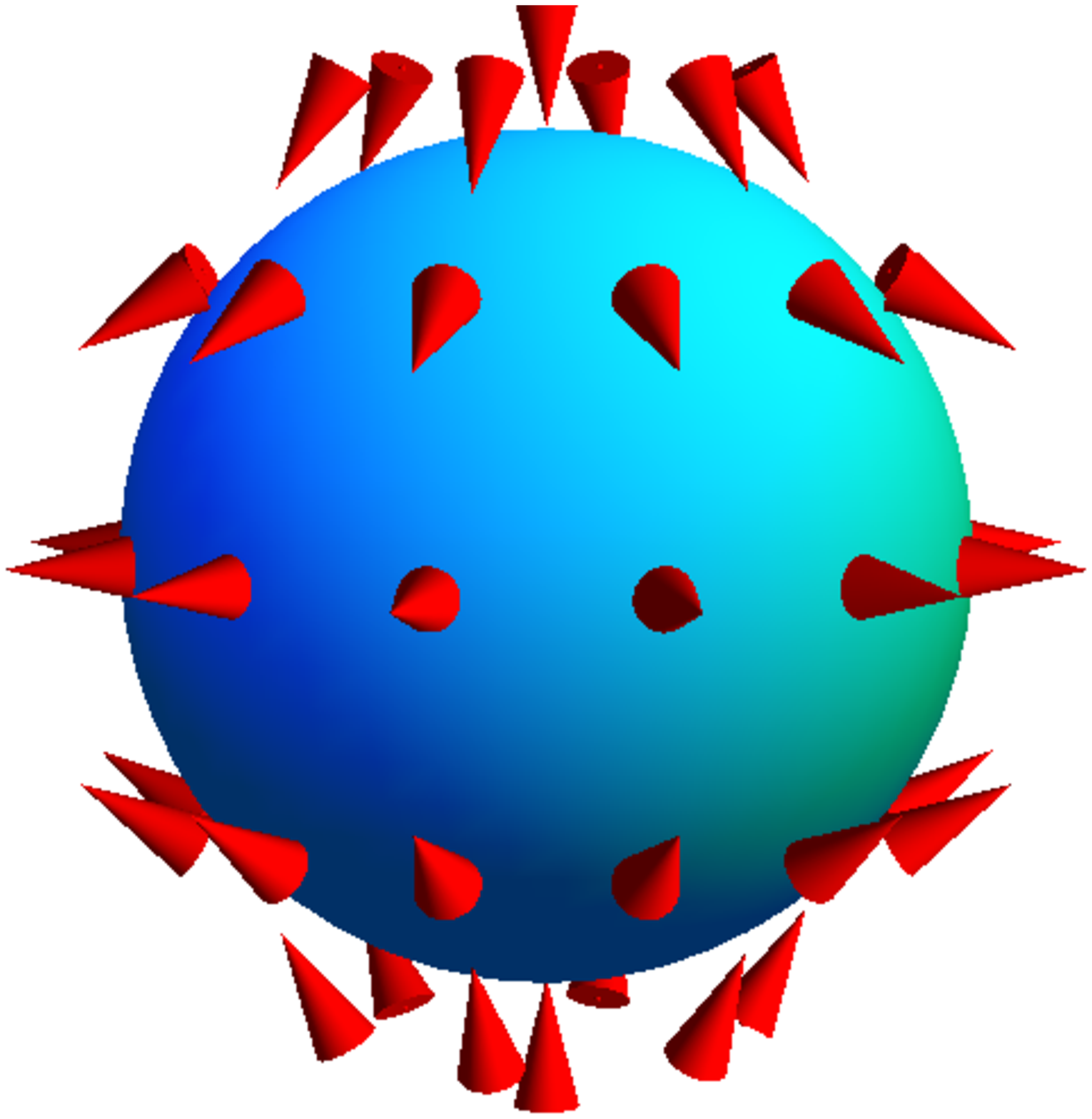} 
\caption{(Color online) Six configurations of a Bloch point with skyrmion number $q=+1$ (top row) and $-1$ (bottom row). Hedgehog Bloch points (\ref{eq:m-simple}) are shown in the left column. Bloch points in the center and right columns are obtained from hedgehogs by a global rotation of magnetization through $90^{\circ}$ and $180^{\circ}$, respectively.}
\label{fig:BP}
\end{figure}

\begin{figure}
\includegraphics[width=0.45\columnwidth]{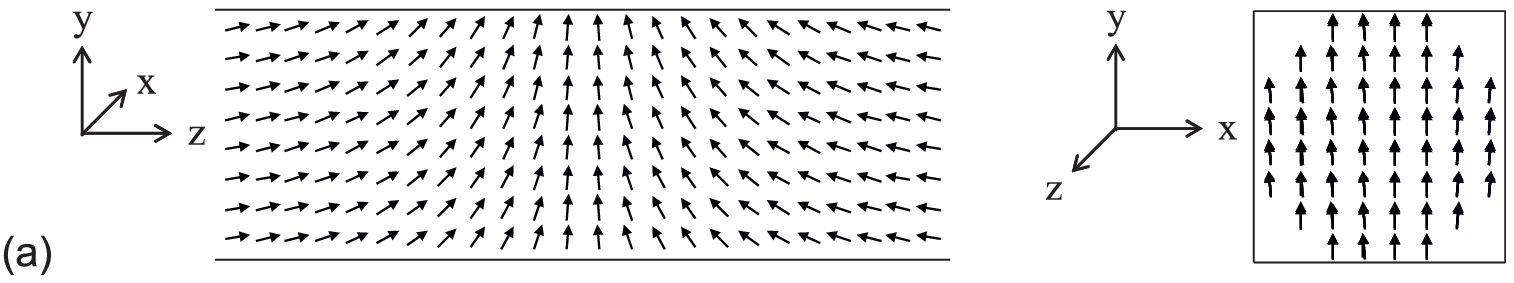}
\hskip 0.5cm
\includegraphics[width=0.45\columnwidth]{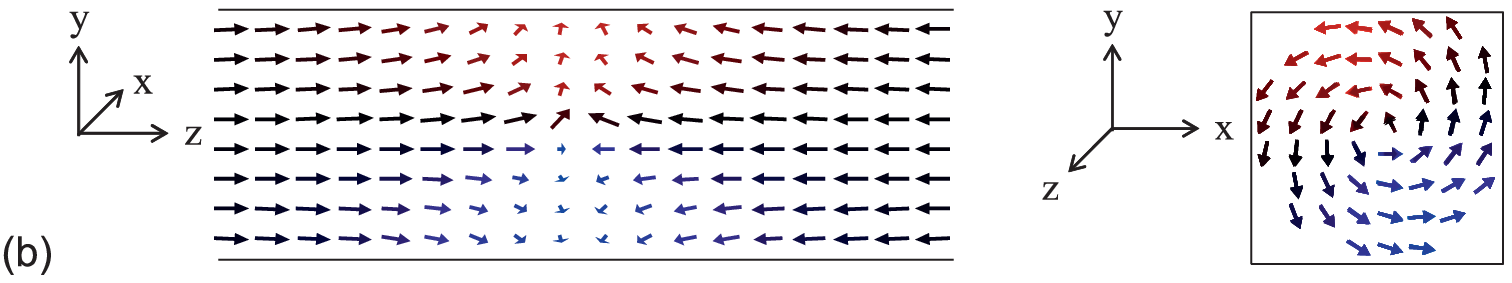} 
\caption{(Color online) (a) The transverse wall in a cylindrical nanowire of radius 10 (nm). (b) The Bloch point wall in a cylindrical nanowire of radius 30 (nm). The left (right) figures are snapshops of magnetization at the center of the cylinder from the negative x (positive z) direction.}
\label{fig:TW_and_BPW}
\end{figure}

Topological defects---nonlinear excitations made stable by their nontrivial topology---are ubiquitous in magnetism. They play an important role in the dynamics of magnets, mediating transitions between low-energy states. Defects come in a variety of dimensions and shapes. In a thin ferromagnetic film, where shape anisotropy keeps magnetization in the plane of the film, point-like topological defects are vortices. Away from the core, magnetization stays in the easy plane, winding clockwise or counterclockwise once as the core is encircled. At the core of a vortex, an area several nanometers across, magnetization points normal to the easy plane. The core endows the magnetization with a nonzero skyrmion density, which is ultimately responsible for a gyrotropic force that dominates vortex dynamics. 

An example of a point-like defect in three dimensions is a Bloch point (BP). \cite{Feldtkeller1965, Doring1968, Guslienko2004, Guslienko2006, Pylypovskyi2012, Lebecki2012} The Fig.~\ref{fig:BP} shows six possible configurations of a BP. In the simplest case of a diverging BP (hedgehog), magnetization in its vicinity has a spherically symmetric distribution, 
\begin{equation}
\mathbf M(\mathbf r) = \pm \frac{M_s \mathbf r}{r}, 
\label{eq:m-simple}
\end{equation}
where $M_s$ is magnetization length, with a typical value of $8\times 10^5$ A/m (permalloy). More generally, a BP is stable if the unit vector of magnetization $\hat{\mathbf m}(\mathbf r) \equiv \mathbf M(\mathbf r)/M_s$ has a nonzero winding number of the second homotopy group $\pi_2(S^2)$,\cite{Volovik1987} also known as the skyrmion charge: 
\begin{equation}
q = \frac{1}{8\pi} \int d A_i \, 
	\epsilon_{ijk} \, \hat{\mathbf m} \cdot \partial_j \hat{\mathbf m} \times \partial_k \hat{\mathbf m},
\label{eq:q}
\end{equation}
where the integral is over a closed surface surrounding the BP. For a divering BP (\ref{eq:m-simple}), $q = \pm 1$. 

A skyrmion number $q$ can also be defined for a vortex on a two-dimensional surface. The skyrmion density $\epsilon_{ijk} \, \hat{\mathbf m} \cdot \partial_j \hat{\mathbf m} \times \partial_k \hat{\mathbf m}/8\pi$ is nonzero only if the unit vector of magnetization $\hat{\mathbf m}$ is non-coplanar. Therefore, the skyrmion charge of a vortex is concentrated near its core, where magnetization comes out of the easy plane. The skyrmion charge of a vortex is half-integer, $q = \pm 1/2$. This difference between a BP and a vortex can be understood intuitively by noting that magnetization $\hat{\mathbf m}$ around a BP acquires every direction on the unit sphere, whereas magnetization near a vortex core only covers half of it, e.g., the northern hemisphere. The skyrmion number of a vortex on the surface of a magnetic sample is related to its ``core polarization,'' i.e., the direction of magnetization normal to the surface. For a vortex, a core magnetized into (out of) the surface has $q = +1/2$ ($-1/2$). For an antivortex, the relation is reversed.

Whereas magnetic vortices have been extensively studied both theoretically and experimentally and are by now well understood, isolated magnetic BPs are elusive and have so far been studied mostly by theorists. (\textcite{Kabanov1989} observed Bloch points on a Bloch line, which in turn is part of a Bloch domain wall.) These objects are fascinating as well as important. For instance, the reversal of core magnetization in a vortex changes its skyrmion number by $\pm 1$. One might suspect that the difference is taken by an object with unit skyrmion charge such as a Bloch point. Indeed, numerical studies have demonstrated the involvement of a Bloch point in a core reversal process. \cite{Thiaville2003, Hertel2006} Hedgehogs also create nontrivial magnetization dynamics. \textcite{Hertel2004} reported an observation of ``magnetic drops'' behind a moving Bloch point. In addition, \textcite{Malozemoff1979} studied Bloch points in the applied context of bubble memories. 

Another peculiar feature of a magnetic BP is the absence of a characteristic length scale below the exchange length, where energy is dominated by the scale-invariant exchange interaction. Whereas the size of a vortex core is determined by a competition between short-exchange force and shape anisotropy due to dipolar interactions, a hedgehog solution (\ref{eq:m-simple}) is scale-invariant and thus sensitive to both short and long-wavelength physics. General BPs are also scale-invariant in a length scale shorter than the exchange length. We may expect that the center of a BP, where spatial variation of magnetization is particularly strong, can be easily pinned by lattice imperfections and even by the atomic lattice itself! In this paper we show that this is indeed the case: the periodic potential of an atomic lattice may create a substantial energy landscape for a Bloch point. 

We present a theoretical study of a Bloch point that is part of a domain wall in a ferromagnetic nanowire with a solid circular cross section (a solid cylinder). Thanks to shape anisotropy of dipolar interactions, such a wire has two ferromagnetic ground states with uniform magnetization along the axis of the wire. A domain wall separating two ground states can be of different types, depending on the radius $R$ of the wire. \cite{Wieser2004, Wieser2010, Usov2007, Yan2010}. For a small radius, not exceeding a few exchange lengths, a ``transverse'' wall is formed, Fig.~\ref{fig:TW_and_BPW}(a). It consists of two surface defects, a vortex and an antivortex with skyrmion numbers $q = \pm 1/2$ each (negative for a head-to-head domain wall). For a larger radius, a ``vortex'' wall is formed, Fig.~\ref{fig:TW_and_BPW} (b). Contrary to its name, it contains a Bloch point, rather than a vortex, so we shall refer to it as a Bloch-point wall (BPW). The skyrmion charge of the BP is $\pm 1$ (negative for a head-to-head domain wall). 

The domain-wall topology of magnetization in a magnetic wire provides an easy way to apply a force to a Bloch point: a uniform magnetic field $\mathbf H_0$ parallel to the axis of the wire pushes the domain wall with a force $\mu_0 Q_m \mathbf H_0$, where $\mu_0$ is the magnetic constant $4 \pi \times 10^{-7}$ H/m, $R$ is the radius of the wire, and $Q_m = \pm 2 \pi R^2 M_s$ is the magnetic charge of the domain wall (positive for a head-to-head domain wall). If the Bloch point is pinned, the wall is unable to move until the applied field is sufficiently strong to overcome the pinning force. 

\begin{figure}
\includegraphics[width=0.2\columnwidth]{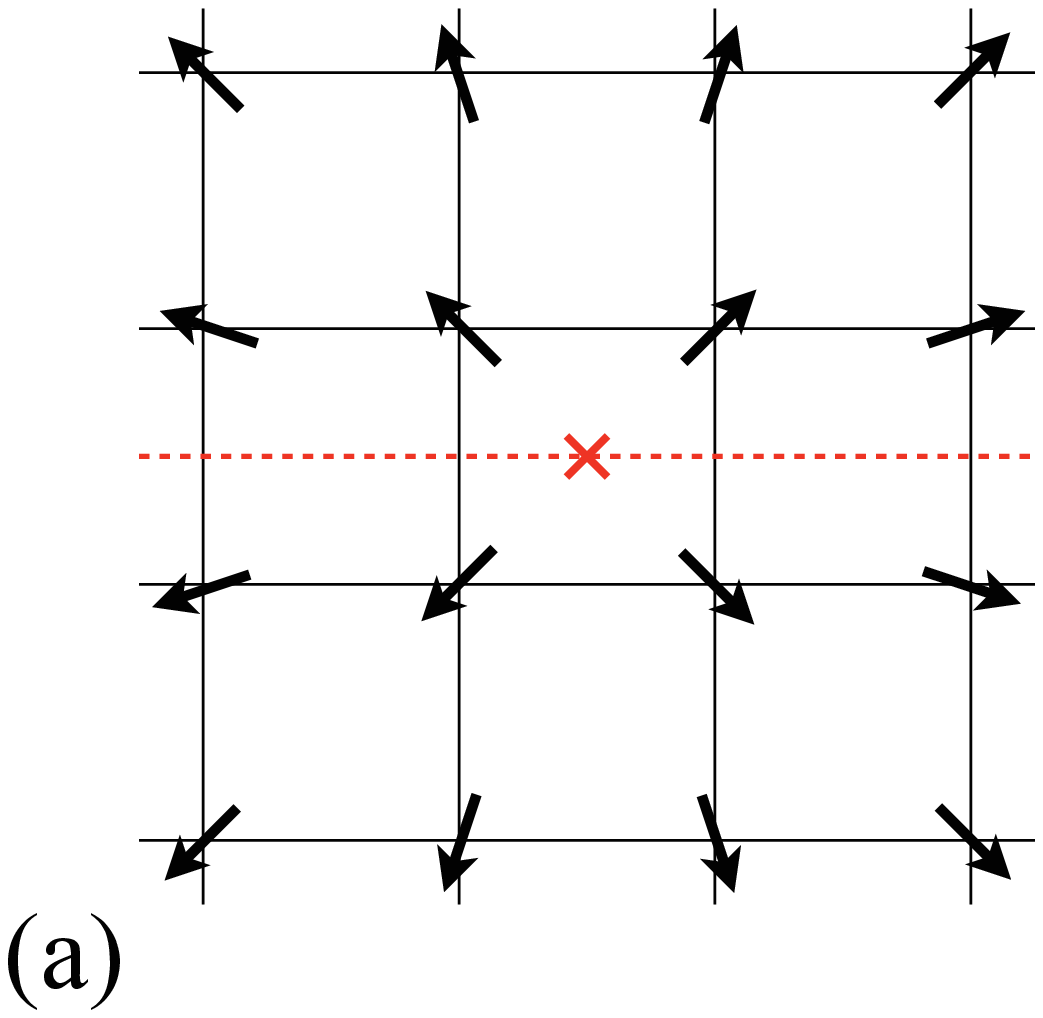}
\hskip 2cm
\includegraphics[width=0.2\columnwidth]{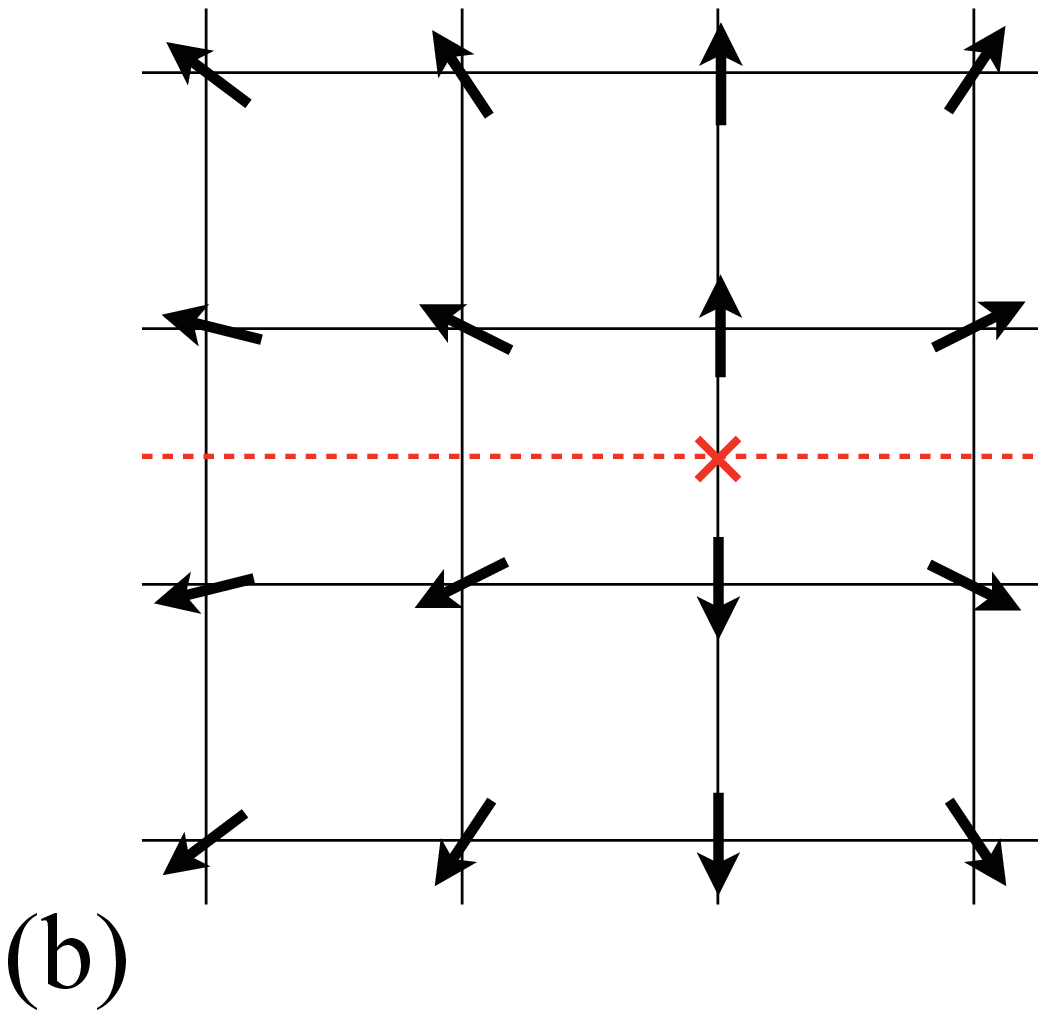} 
\caption{(Color online) Magnetization configurations with two different BP positions (cross): (a) at the center of a cubic cell and (b) at the center of a cell's face.}
\label{fig:BP_in_the_cell}
\end{figure}

The pinning of a Bloch point in an atomic lattice has a simple origin. Picture a BP moving through a simple cubic lattice of magnetic atoms parallel to one of the cubic axes, Fig.~\ref{fig:BP_in_the_cell}. The BP's energy is lowest when it is at the center of a cubic cell ($z = 0$), farthest from the magnetic dipoles in its corners. It is highest (along that trajectory) when the BP is at the center of a face ($z = a/2$). A simple model of the Bloch point's energy on this straight line is $U(z) = -U_0 \cos{(2\pi z/a)}$, where $a$ is the lattice constant. The amplitude $U_0$ can be estimated as follows. The primary source of the pinning potential is the exchange energy $E_\mathrm{ex} = A \int d^3r \, \partial_i \hat{\mathbf m} \cdot \partial_i \hat{\mathbf m}$, where $A$ is the exchange constant with the dimension of energy per unit length. We thus expect that $U_0 = c A a$, where $c$ is a numerical constant of order 1. This gives the maximal pinning force $F_p = 2 \pi c A$ that is independent of the lattice constant $a$.  The critical field for depinning a domain wall with a Bloch point in a magnetic wire of radius $R$ is then 
\begin{equation}
H_c = \frac{cA}{\mu_0 M_s R^2} = \frac{cM_s}{2} \frac{\ell_\mathrm{ex}^2}{R^2}, 
\end{equation}
where $\ell_\mathrm{ex} = \sqrt{2 A/\mu_0 M^2_s}$ is the exchange length. For a typical ferromagnet (permalloy, $A = 1.3 \times 10^{-11}$ J/m, $M_s = 8 \times 10^5$ A/m, $\ell_\mathrm{ex}$ = 5.7 nm), we obtain a pinning force of order $10^{-11}$ N. A permalloy wire with radius $R = 50$ nm would have a sizable depinning field $H_c \approx 60$ Oe assuming $c \approx 1$. The numerical constant $c$ of course cannot be determined by dimensional analysis and will be derived using a variational model of a Bloch point below. 

The paper is organized as follows. In Sec. \ref{sec:theory}, we compute the lattice potential of a magnetic BP and the associated critical field for a Bloch-point domain wall. In Sec. \ref{sec:simulation}, we compare the computed value against numerical simulations using the micromagnetic simulator OOMMF.\cite{Donahue1999} We discuss our results in Sec. \ref{sec:conclusion}.

\section{Theory for the pinning critical field}
\label{sec:theory}

\begin{figure}
\includegraphics[width=0.2\columnwidth]{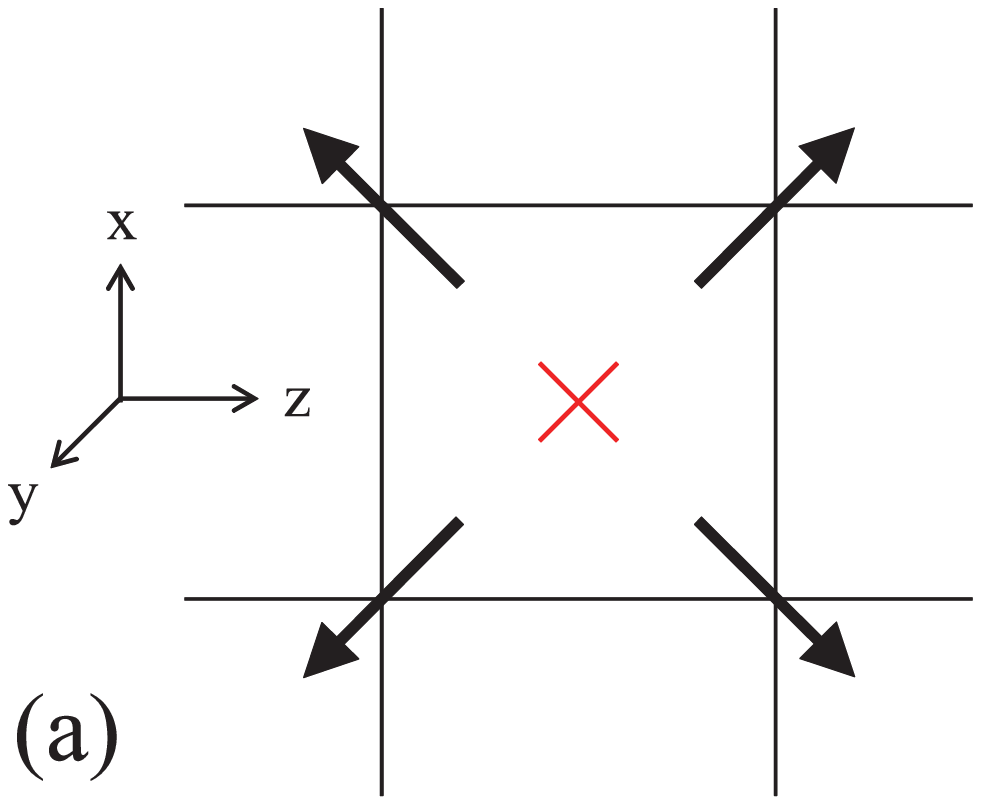} 
\hskip 0.5cm
\includegraphics[width=0.2\columnwidth]{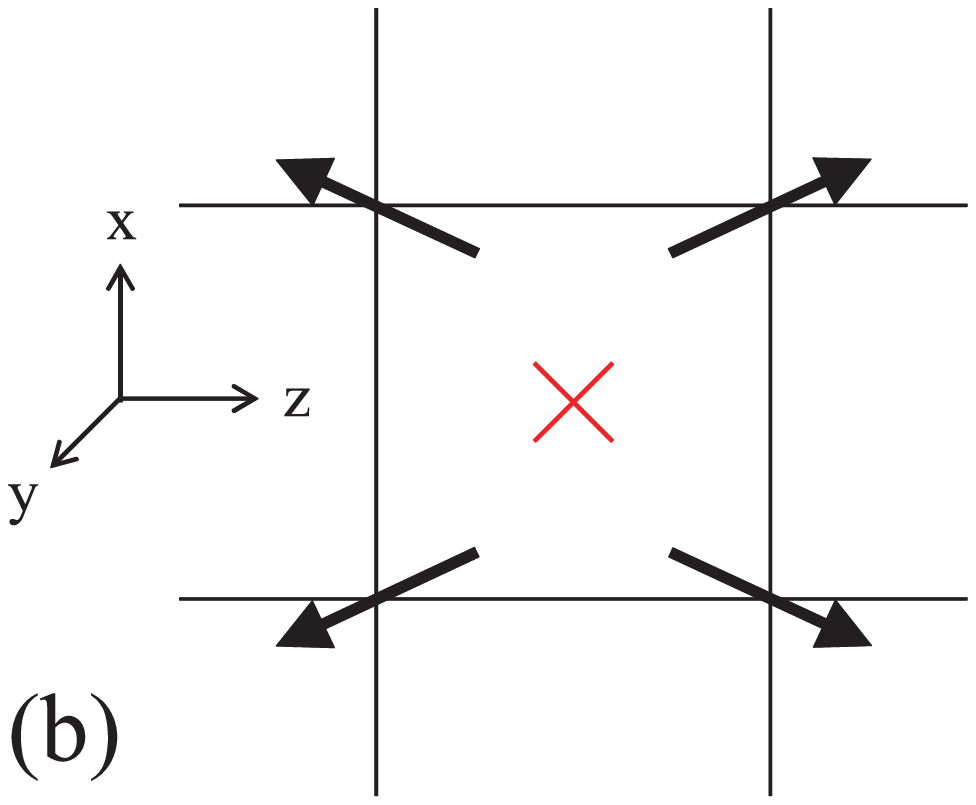} 
\hskip 0.5cm
\includegraphics[width=0.2\columnwidth]{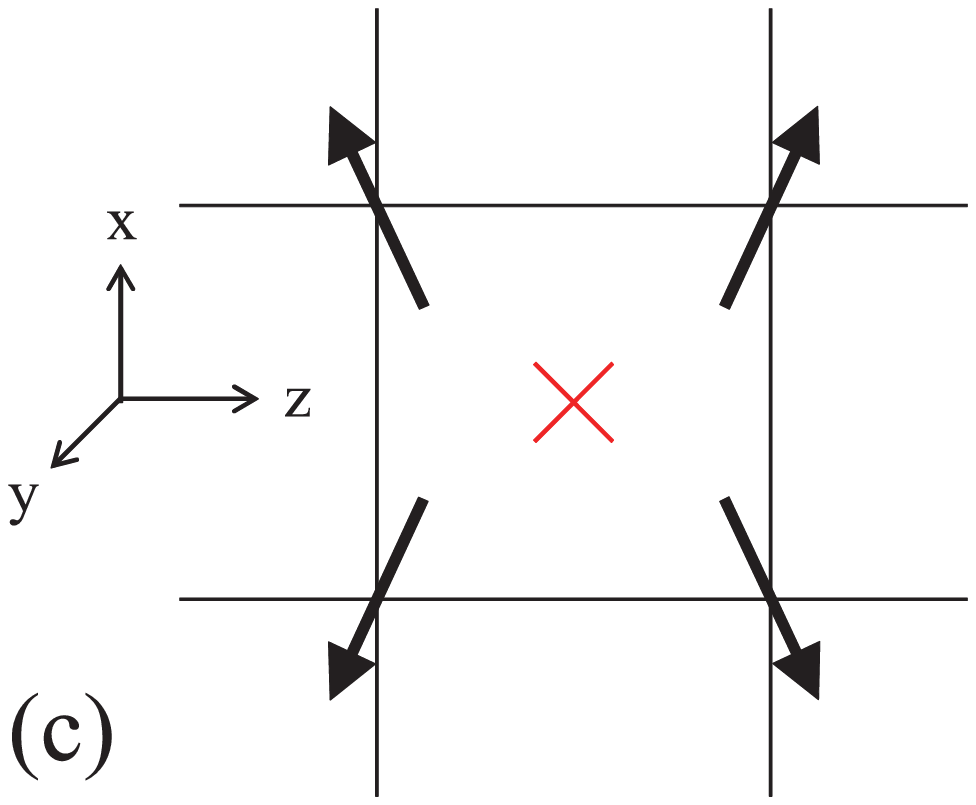} 
\caption{(Color online) The effect of the shape factor $s$ on the configuration of magnetic moments closest to a BP (cross). (a) $s = 1$. (b) $s < 1$. (c) $s > 1$.}
\label{fig:s}
\end{figure}

The energy of a ferromagnet in a continuum model is given by
\begin{equation}
E = \int d^3r \left( 
	A \, \partial_i \hat{\mathbf m} \cdot \partial_i \hat{\mathbf m} 
		- \frac{\mu_0 M^2_s}{2} \hat{\mathbf m} \cdot \mathbf h_D - K m_z^2 
	\right), 
\label{eq:E-continuum}
\end{equation}
where $\hat{\mathbf m} \equiv \mathbf M/M_s$ is the unit magnetization vector, $A$ is the exchange constant, $M_s$ is the saturation magnetization, $K$ is an easy-axis anisotropy constant, and $\mathbf h_D$ is a dimensionless demagnetizing field satisfying $\nabla \cdot \mathbf h_D = - \nabla \cdot \hat{\mathbf m}$ and $\nabla \times \mathbf h_D = 0$. The coupling constants $A$, $\mu_0 M^2_s$, and $K$ have dimensions of energy/length, energy/length$^3$, and energy/length$^3$, respectively. They can be used to construct an intrinsic length scale known as the exchange length $\ell_\mathrm{ex} \equiv \sqrt{2 A/\mu_0 M^2_s}$ and a dimensionless ``quality factor'' $Q = 2 K/\mu_0 M^2_s$ expressing the strength of local anisotropy relative to the shape anisotropy of the demagnetizing field $\mathbf h_D$. Since we are interested in a weakly anisotropic system, $Q \ll 1$, we ignore the anisotropy energy for calculation of the mesh potential of BP. 

We have evaluated the energy of a magnetic BP in a Heisenberg ferromagnet on a simple cubic lattice with nearest-neighbor exchange coupling $J = 2 Aa$, which gives the same exchange energy for slow spatial variations as does the continuum version, Eq.~(\ref{eq:E-continuum}). The lattice Hamiltonian is
\begin{equation}
E = - 2 A a \sum_{\langle ij \rangle} \hat{\mathbf m}_i \cdot \hat{\mathbf m}_j 
	+ \frac{A a^3}{4 \pi \ell_\mathrm{ex}^2} \sum_{i \ne j} 
		\frac{\hat{\mathbf m}_i \cdot \hat{\mathbf m}_j 
		- 3 (\hat{\mathbf m}_i \cdot \hat{\mathbf e}_{ij}) (\hat{\mathbf m}_j \cdot \hat{\mathbf e}_{ij})}
		{r^3_{ij}}
\label{eq:ex-dip}
\end{equation}
where $\hat{\mathbf e}_{ij}$ is the unit vector pointing from the site $i$ to the site $j$ and $r_{ij}$ is the distance between two sites in units of the mesh size $a$. Spins are located at half-integer lattice points $\mathbf r_i = (i_1 + 1/2, i_2 + 1/2, i_3 + 1/2)$, where $i_1$, $i_2$, and $i_3$ are integers. We used the following Ansatz for magnetization in cylindrical coordinates $\mathbf r = (\rho, \phi, z)$, $z$ being the wire axis: 
\begin{eqnarray}
m_{\rho} 	&=& \frac{s \rho \sin\phi_0}{\sqrt{s^2 {\rho}^2 + (z-\zbp)^2}},  
\nonumber \\
m_{\phi} 	&=& \frac{s \rho \cos\phi_0}{\sqrt{s^2 {\rho}^2 + (z-\zbp)^2}}  
\label{eqn:ansatz} \\
m_{z}		&=& \frac{z-\zbp}{ \sqrt{s^2 {\rho}^2 + (z-\zbp)^2}} 
\nonumber
\end{eqnarray}
where $(0, 0, \zbp)$ is the location of BP, $\phi_0$ is the deviation magnetization projected into the $xy$ plane from the azimuthal direction $\hat{\bm \phi}$, and $s$ is a phenomenological shape factor expressing the deformation of a BP inside a cylindrical wire. Please see the Fig.~\ref{fig:s} for the possible configurations of magnetization depending on the shape factor. The undeformed BP is spherically symmetric, so the shape factor $s=1$. BPs with $s>1$ ($s<1$) can be obtained by expanding (shrinking) a spherical solution along the z-direction. In other words, when $s >1$ $(s < 1)$, $m_z$ is smaller (larger) than $m_z$ for $s = 1$. The shape factor was determined from numerical simulations by fitting magnetization of a static BP to Ansatz (\ref{eqn:ansatz}) in the vicinity of the singularity. 

\begin{figure}
\includegraphics[scale=0.7]{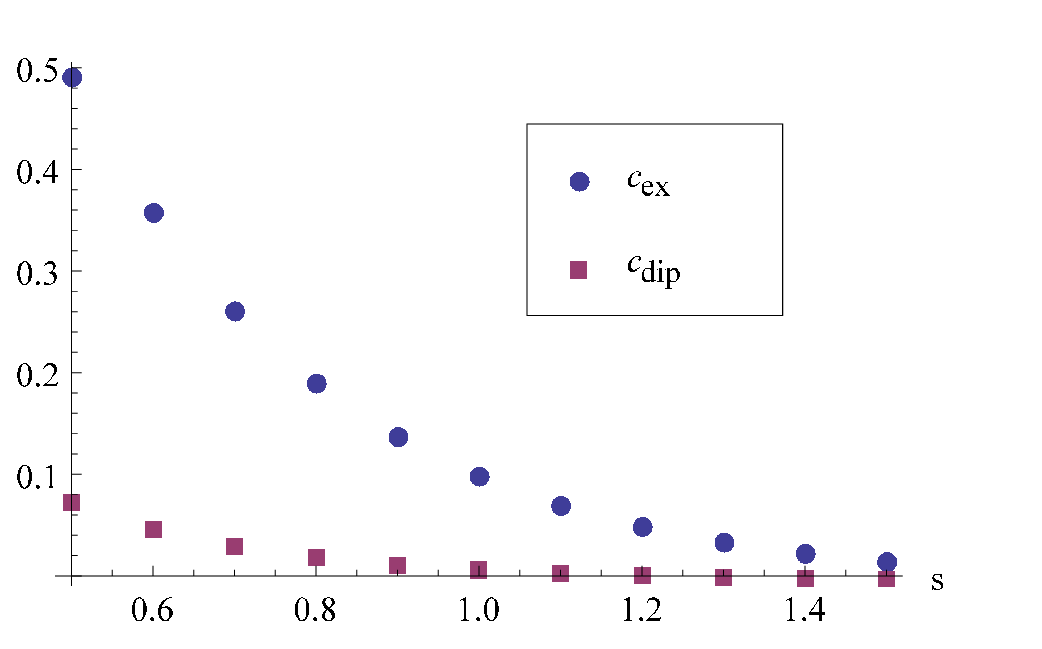}
\caption{(Color online) Dependence of $c_\mathrm{ex}$ and $c_\mathrm{dip}$ on the shape factor $s$. }
\label{fig:c_vs_s}
\end{figure}

Our analysis, presented in detail in Appendix \ref{sec:appendix}, shows that the periodic potential experienced by the BP in the lattice is dominated by the first harmonic: 
\begin{equation}
E(\zbp) \approx - A a(c_\mathrm{ex} - c_\mathrm{dip} a^2/\ell_\mathrm{ex}^2)\cos{(2\pi\zbp/a)}, 
\label{eq:E-zbp}
\end{equation}
where $c_\mathrm{ex}$ and $c_\mathrm{dip}$ are positive constants of order of 1 reflecting the contributions of exchange and dipolar interactions. Figure~\ref{fig:c_vs_s} shows the dependence of $c_\mathrm{ex}$ and $c_\mathrm{dip}$ on the shape factor $s$. In the case of the atomic lattice, $a \ll \ell_\mathrm{ex}$ and the dipolar term contributes relatively little. However, in our numerical simulations the lattice constant $a$ was comparable to the exchange length $\ell_\mathrm{ex}$. We therefore kept the dipolar term in Eq.~(\ref{eq:E-zbp}) for comparison with the numerics. 

The pinning force created by the periodic potential is
\begin{equation}
F_p(\zbp) = - E'(\zbp) \approx 2 \pi A (c_\mathrm{ex} - c_\mathrm{dip} a^2/\ell_\mathrm{ex}^2) 
	\sin{(2\pi\zbp/a)}.
\end{equation}
It matches the external force of the applied field, $F_Z = \mu_0 Q_m H$, where $\mu_0$ is the magnetic constant $4 \pi \times 10^{-7}$ H/m, $R$ is the radius of the wire, and $Q_m = 2\pi M_s R^2$ is the wall's magnetic charge, until $F_p$ reaches its maximum value at $\zbp = a/4$. This gives a critical field for depinning, 
\begin{equation}
H_c = (c_\mathrm{ex} - c_\mathrm{dip} a^2/\ell_\mathrm{ex}^2) \frac{M_s}{2} \frac{l_{ex}^2}{R^2} .
\end{equation}

\section{Numerical simulation}
\label{sec:simulation}
\begin{figure}
\includegraphics[width=0.4\columnwidth]{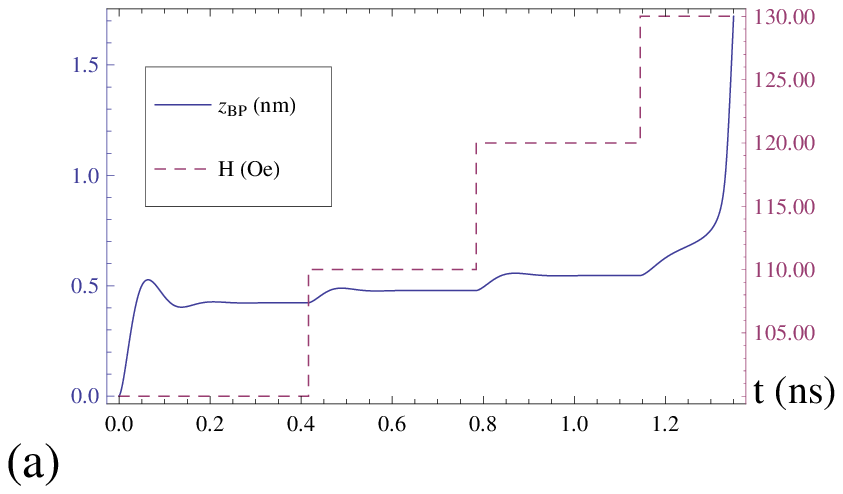} 
\hskip 1cm
\includegraphics[width=0.4\columnwidth]{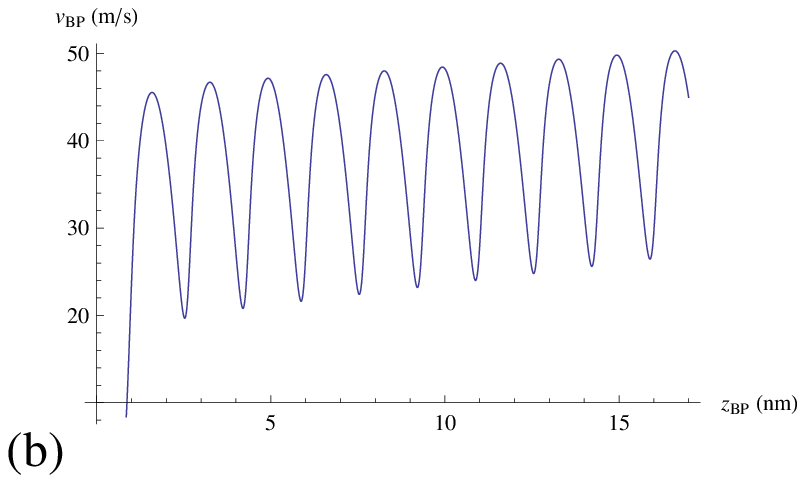} 
\caption{(Color online) (a) The position of the Bloch point and external field as a function of time. (b) The velocity of the Bloch point as a function of its position. The lattice constant is 1.7 nm.}
\label{fig:mz_vs_t}
\end{figure}

To test our idea, we performed numerical simulations using the micromagnetic package OOMMF.\cite{Donahue1999} Its three-dimensional numerical solver \texttt{Oxs} treats magnetization as a discrete variable, a unit vector $\hat{\mathbf m}_i$ defined on sites $i$ of a simple cubic lattice. The exchange energy is computed as a sum of scalar products $- \hat{\mathbf m}_i \cdot \hat{\mathbf m}_j$ for pairs of nearest-neighbor sites $i$ and $j$, precisely as in the Heisenberg ferromagnet on a cubic lattice. See Refs.~\onlinecite{Donahue1999, Donahue1997, Porter2001, Donahue2004} for details. 

The simulated sample had the following material parameters:\cite{Usov2007} exchange constant $A = 1.0 \times 10^{-11}$ J/m, saturation magnetization $M_s = 10^6$ A/m, easy-axis anisotropy $K = 10^4$ J/m$^3$,  gyromagnetic ratio $\gamma_0 = 2.21 \times 10^5$ m/(A s), and a damping constant $\alpha = 0.3$. These material parameters are slightly modified from permalloy to reduce simulation time. The exchange length for this material is $\ell_\mathrm{ex} = 4.0$ nm and the quality factor is $Q = 0.016$. We added a small easy-axis anisotropy to stabilize a domain wall with a Bloch point.\cite{Wieser2004} Since $Q \ll 1$, we did not include the anisotropy energy in the calculation of the mesh potential. We used two different sample geometries: (i) radius $R = 10$ nm and length $L = 200$ nm; (ii) radius $R = 15$ nm and length $L = 300$ nm. The stable configurations of BPWs are obtained by conjugate gradient minimizer in OOMMF \cite{Donahue1999} starting from the undeformed circulating BPWs ($s$ = 1, $\phi_0 = 0$ in Ansatz (\ref{eqn:ansatz})). The lattice constants are chosen so that a BP sits at the center of a cubic cell.

To determine the critical field, we applied the external field increasingly starting at 100 Oe and 30 Oe and increasing it in steps of 10 Oe and 5 Oe for the narrow and wide wires, respectively. Fig.~\ref{fig:mz_vs_t} shows the position of the Bloch point $\zbp$ as a function of time for the narrow wire with a lattice constant $a$ = 1.7 nm. 

\begin{figure}
\includegraphics[scale=0.7]{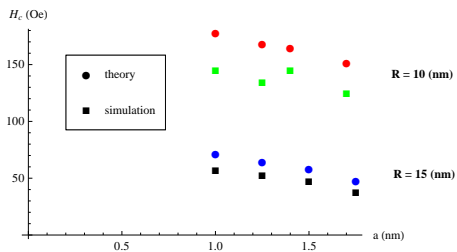}
\caption{(Color online) Theoretical and simulated critical fields}
\label{fig:result}
\end{figure}

\begin{figure}
\includegraphics[scale=1]{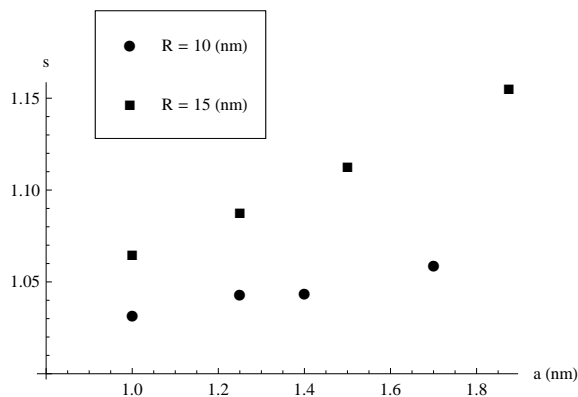}
\caption{(Color online) The dependence of the shape factor $s$ on the lattice constant $a$}
\label{fig:a_to_s}
\end{figure}

The results are summarized in Fig.~\ref{fig:result}. It can be seen that the depinning field $H_c$ does not extrapolate to zero as the size of a unit cell $a$ in the simulation is reduced. This behavior of a Bloch-point domain wall is strikingly different from the typical behavior of domain walls in  numerical simulations. Ordinarily, the effects of lattice discretization become unnoticeable when the mesh size is reduced below half the exchange length $\ell_\mathrm{ex}$. The finest mesh size in our simulations, $a = 1$ nm, is much smaller than $\ell_\mathrm{ex} = 4.0$ nm, yet the lattice effects do not subside. Furthermore, the critical field \emph{increases} as the mesh size decreases. The main reason for this effect comes from the dependence of the shape factor on the mesh size. As the mesh size decreases, the shape factor decreases, which results in increase in $c_{ex}$ (see Fig.~\ref{fig:a_to_s}). Magnetization around BP becomes more isotropic as the mesh size decreases.

The mesh potential also affects dynamics of a BPW under the external field above the critical field. The velocity of BPW oscillates and this oscillation is periodic in the position of BPW, which means that the mesh potential for BP is responsible for the oscillation. See (b) in the figure~\ref{fig:mz_vs_t} for an example. The slight increase of the average velocity as a function of time can be attributed to the attractive force between the volume magnetic charge of BPW and the surface magnetic charge at the end of the cylinder.

\section{Discussion}
\label{sec:conclusion}

Topological defects in magnets can be pinned by a periodic potential of the underlying magnetic lattice. When lattice discretization is employed to simulate a continuous magnetic medium, this periodic potential is an artifact that can be removed by choosing a sufficiently fine mesh. For a domain wall, the lattice period should be small compared to the wall width; for a vortex, smaller than its core. A magnetic Bloch point turns out to be exceptional in this regard because it has no characteristic length scale. As a result of that, the amplitude of the periodic pinning potential of the lattice is linearly proportional to the lattice constant. Thus the maximal pinning force is independent of the lattice constant and is of the order of the exchange constant, $A = 1.3 \times 10^{-11}$ N in permalloy. 

In this paper we have estimated the strength of the periodic pinning potential experienced by a magnetic Bloch point in a simple cubic lattice. The estimate is in good agreement with micromagnetic simulations, which yield a finite pinning field for a domain wall containing a Bloch point, even when the unit cell is much smaller than the exchange length. We expect that Bloch points are strongly pinned by the atomic lattice in ferromagnets. The strength of pinning should be reduced by thermal and quantum fluctuations, which can create a core where magnetization vanishes. Numerical estimates\cite{Elias2011, Lebecki2012} show that the core size induced by thermal fluctuations is small, comparable to the lattice spacing at room temperature. 

Previous authors have shown the velocity of a BPW to be proportional to the applied field. \cite{Wieser2004} This result is valid in the high-field limit ($H \gg H_c$). The linear relation does not hold just above the pinning threshold, where the pinning force is comparable to the force from the external field.

\begin{acknowledgments}
The authors are grateful to Christopher Mogni for useful discussions. This work was supported in part by the U.S. National Science Foundation under Grant No. DMR-1104753.
\end{acknowledgments}

\appendix
\section{Mesh potential calculation}
\label{sec:appendix}

The exchange and dipolar energies of a ferromagnet (\ref{eq:ex-dip}) in the presence of a Bloch point at $(0,0,\zbp)$ are 
\begin{eqnarray}
E_{\mathrm{ex}}(\zbp)	&=& 
- 2 A a \sum_{\langle ij \rangle}  \hat{\mathbf m}_i \cdot \hat{\mathbf m}_j  
 \nonumber \\
E_{\mathrm{dip}}(\zbp)	&=&  \frac{Aa^3}{4\pi \ell_{\mathrm{ex}}^2} \sum_{i \ne j}  \frac{\hat{\mathbf m}_i \cdot \hat{\mathbf m}_j - 3 (\hat{\mathbf m}_i \cdot \hat{\mathbf e}_{ij})(\hat{\mathbf e}_{ij} \cdot \hat{\mathbf m}_j)}{|{\mathbf r}_i - {\mathbf r}_j|^3},
\label{eqn:energy}
\end{eqnarray}
with $\hat{\mathbf m}_i$ given by Ansatz (\ref{eqn:ansatz}) where $A$ is the exchange constant, $\ell_\mathrm{ex} \equiv \sqrt{2 A/\mu_0 M^2_s}$ is the exchange length, $\mu_0$ is the magnetic constant, and $M_s$ is the saturation magnetization. Being periodic functions of $\zbp$ with a lattice period $a$, these expressions can be written as Fourier series: 
\begin{eqnarray}
E_{\mathrm{ex}}(\zbp) - E_{\mathrm{ex}}(0) &=& 
	- A a \sum_{n=1}^{\infty} c_{\mathrm{ex},n} \cos(2 \pi n \zbp),
\nonumber\\
E_{\mathrm{dip}}(\zbp) - E_{\mathrm{dip}}(0) &=& 
	\frac{Aa^3}{\ell_{\mathrm{ex}}^2} \sum_{n=1}^{\infty} c_{\mathrm{dip},n} \cos(2 \pi n \zbp).
\end{eqnarray}
The subtraction of the $\zbp=0$ term improves convergence of the energy in a finite lattice. The Fourier amplitudes $c_{\mathrm{ex},n}$ and $c_{\mathrm{dip},n}$ depend on the shape factor $s$; $c_\mathrm{dip,n}$ also depends on the angle $\phi_0$. The overall sign convention is such that the first harmonics $c_\mathrm{ex,1}$ and $c_\mathrm{dip,1}$ are positive. The shape factor $s$ and the angle $\phi_0$ were obtained by fitting magnetization of an OOMMF solution at the eight mesh points closest to a Bloch point at $\zbp=0$ to Ansatz (\ref{eqn:ansatz}).

The Fourier amplitudes $c_\mathrm{ex,n}$ can be computed analytically by applying Poisson's formula to the index $i_3$ along the z-axis, 
\[
\sum_{i_3=-\infty}^{\infty} f(z + i_3) = \sum_{n=-\infty}^{\infty}e^{-2 \pi i n z} \int dz' e^{2 \pi i n z'} f(z').
\]
We thus obtain
\begin{eqnarray}
c_{\mathrm{ex},n}	&=& 2 \sum_{i_1, i_2 = -\infty}^{\infty} \int dz \cos (2 \pi n z) \nonumber \\
			&& \times \left[ 4 (-1)^n \left( \frac { s \mathbf{x}_i + z \hat{\mathbf z} }{ |s \mathbf{x}_i + z \hat{\mathbf z}| } \cdot \frac { s \mathbf{x}_i - s \hat{\mathbf x} + z \hat{\mathbf z} }{ |s \mathbf{x}_i - s \hat{\mathbf x} + z \hat{\mathbf z}| } - 1 \right) + 2 \left( \frac { s \mathbf{x}_i + (z + 1/2) \hat{\mathbf z} }{ |s \mathbf{x}_i + (z + 1/2) \hat{\mathbf z}| } \cdot \frac { s \mathbf{x}_i + (z - 1/2) \hat{\mathbf z} }{ |s \mathbf{x}_i + (z - 1/2) \hat{\mathbf z}| } -1 \right) \right].
\end{eqnarray}
Here $\mathbf{x}_i = (i_1 + 1/2, i_2 + 1/2, 0)$ is the projection of $\mathbf{r}_i$ onto the xy plane and $\phi_0 = \pi/2$ is used. This expression is dominated by the $n=1$ term. In practice it suffices to use only $n=1$ and $i_1, i_2 = -1, 0, 1$. $c_\mathrm{ex}$ used in the main part of the paper is $c_\mathrm{ex,1}$ calculated with $i_1, i_2 = -1, 0, 1$.

The Fourier amplitudes $c_\mathrm{ex,1}$ and $c_\mathrm{dip,1}$ have also been evaluated numerically. After setting the maximum value on $I \equiv \max(|i_1|, |i_2|, |i_3|)$ in Eq.(~\ref{eqn:energy}), we computed the dependence of $E_\mathrm{ex}$ and $E_\mathrm{dip}$ on $\zbp$ by increasing $\zbp$ from 0 to 1 in steps of 0.1. We then extracted $c_\mathrm{ex,1}$ and $c_\mathrm{dip,1}$ by a Fourier transformation. These coefficients converge rapidly as $I$ increases and $I = 10$ provides good accuracy. $c_\mathrm{dip}$ used in the main part of the paper is numerically calculated $c_\mathrm{dip,1}$.

\bibliography{MyCollection}

\end{document}